\documentclass[prd,aps,twocolumn,amsmath,showpacs,nofootinbib]{revtex4}
\usepackage{amsmath}
\usepackage{amssymb}
\usepackage{amsfonts}
\usepackage{graphicx,bm}
\usepackage{dcolumn}
\usepackage{color,amsxtra}
\usepackage{epsf}
\usepackage{enumerate}
\usepackage{hhline}
\usepackage{array}
\usepackage{tabularx}
\newcommand{\be}{\begin{equation}}
\newcommand{\ee}{\end{equation}}
\newcommand{\bea}{\begin{eqnarray}}
\newcommand{\eea}{\end{eqnarray}}
\newcommand{\beaa}{\begin{eqnarray*}}
\newcommand{\eeaa}{\end{eqnarray*}}

\newcommand{\e}{\mathrm{e}}

\newcommand{\abs}[1]{\vert{#1}\vert}
\def\be{\begin{equation}}
\def\ee{\end{equation}}
\def\bea{\begin{eqnarray}}
\def\eea{\end{eqnarray}}

\def\e{\mathrm{e}}
\begin{document}
%\draft

%\tolerance=5000

%%%%%%%%%%%%%%%%%%%%%
%  Title
%%%%%%%%%%%%%%%%%%%%%
\title{Effective $F(T)$ gravity from the higher-dimensional Kaluza-Klein and Randall-Sundrum theories}
\author{Kazuharu Bamba$^{1,}$\footnote{
E-mail address: bamba@kmi.nagoya-u.ac.jp},
Shin'ichi Nojiri$^{1, 2,}$\footnote{E-mail address:
nojiri@phys.nagoya-u.ac.jp} 
and 
Sergei D. Odintsov$^{3, 4, 5,}$\footnote{
E-mail address: odintsov@ieec.uab.es}
}
\affiliation{
$^1$Kobayashi-Maskawa Institute for the Origin of Particles and the
Universe,
Nagoya University, Nagoya 464-8602, Japan\\
$^2$Department of Physics, Nagoya University, Nagoya 464-8602, Japan\\
$^3$Instituci\`{o} Catalana de Recerca i Estudis Avan\c{c}ats (ICREA), 
Barcelona, Spain\\
$^4$Institut de Ciencies de l'Espai (CSIC-IEEC), 
Campus UAB, Facultat de Ciencies, Torre C5-Par-2a pl, E-08193 Bellaterra
(Barcelona), Spain\\ 
$^5$Tomsk State Pedagogical University, Tomsk, Russia
}

%\date{\today}

%%%%%%%%%%%%%%%%%%%%%
%  Abstract
%%%%%%%%%%%%%%%%%%%%%
\begin{abstract}

We explore the four-dimensional effective $F(T)$ gravity with 
$T$ the torsion scalar in teleparallelism  
originating from higher-dimensional space-time theories, 
in particular the Kaluza-Klein (KK) and Randall-Sundrum (RS) theories.  
First, through the KK dimensional reduction from 
the five-dimensional space-time, 
we obtain the four-dimensional effective theory 
of $F(T)$ gravity with its coupling to a scalar field. 
Second, taking the RS type-II model in which there exist 
the five-dimensional Anti-de Sitter (AdS) space-time 
with four-dimensional Friedmann-Lema\^{i}tre-Robertson-Walker 
(FLRW) brane, we find that there will appear the contribution of 
$F(T)$ gravity on the four-dimensional FLRW brane. 
It is demonstrated that 
inflation and the dark energy dominated stage can be realized 
in the KK and RS models, respectively, due to the effect of only the torsion 
in teleparallelism without that of the curvature. 

\end{abstract}
%%%%%%%%%%%%%%%%%%%%%

%----------------------------
%\pacs{
%Keywords:
%}
%\preprint{}
%\hspace{13.0cm}
%----------------------------

\maketitle
%==============================================================================

%%%%%%%%%%%%%%%%%%%%%%%%%%%
%%%  Sec. I
%%%%%%%%%%%%%%%%%%%%%%%%%%%
\section{Introduction}
%\noindent
%%%%%%%%%%%%%%%%%%%%%
%\textit{Introduction}--- 
%%%%%%%%%%%%%%%%%%%%%

The phenomenon of the accelerated expansion of the universe 
has been supported by various observations of 
Supernovae Ia~\cite{SN1}, large scale structure~\cite{LSS} 
including the baryon acoustic oscillations~\cite{Eisenstein:2005su}, 
cosmic microwave background radiation~\cite{WMAP}, 
and weak lensing~\cite{Jain:2003tba}. 
This is one of the most significant problems in modern cosmology. 
Provided that the universe is homogeneous, 
there exist two representative ways of accounting for the current cosmic 
acceleration: The first is to introduce 
``dark energy'', which has negative pressure, 
within general relativity (for recent reviews, see, e.g.,~\cite{R-DE}). 
The second is to modify the gravitational theory 
on large scale. 
As one of the latter approaches, 
``teleparallelism''~\cite{Teleparallelism} 
has recently been drawn much attention. 
The formulations are constructed with 
the Weitzenb\"{o}ck connection, and hence 
the action is described by the torsion scalar $T$, whereas 
in general relativity, the formulations are written with the Levi-Civita connection, and thus the action is represented by 
the scalar curvature $R$. 
It has been illustrated that in $F(T)$ gravity, 
inflation in the early universe~\cite{Inflation-F-F} or 
the late-time cosmic acceleration~\cite{Bengochea:2008gz, Linder:2010py, WY-BGLL-BGL} can be realized. 
Also, it was verified that a non-minimal gravitational coupling of 
a scalar field in teleparallelism can explain 
the current cosmic acceleration~\cite{Geng:2011aj}. 
Various theoretical issues of $F(T)$ gravity have extensively 
been discussed. 
%%%%%%%%

In this Letter, 
we examine the four-dimensional effective $F(T)$ gravity taking its origin from higher-dimensional space-time theories. 
As the first example, we consider the four-dimensional effective $F(T)$ gravity from the Kaluza-Klein (KK) theory~\cite{MKK-A-C-F, Overduin:1998pn, F-M}. 
By setting up the five-dimensional space-time, 
via the KK reduction to the four-dimensional space-time, 
we construct the four-dimensional effective theory, 
which is an $F(T)$ gravity model with the non-minimal 
coupling to a scalar field. 
Next, as the second example, 
we investigate the four-dimensional effective $F(T)$ gravity from 
the Randall-Sundrum (RS)~\cite{Randall:1999ee, Randall:1999vf} theory, 
which originates from a novel KK approach in the brane world 
description~\cite{AHDD-AAHDD}. 
We take the RS type-II model where there are 
the five-dimensional Anti-de Sitter (AdS) space-time and the 
four-dimensional Friedmann-Lema\^{i}tre-Robertson-Walker 
(FLRW) brane. In such a configuration, 
a contribution of $F(T)$ gravity on the FLRW brane will exist. 
It is shown that 
inflation or the dark energy dominated stage can be realized 
only by the effect of the torsion without that of the curvature. 
As a result, it can be interpreted that 
these models may be equivalent to the KK and RS models without gravitational 
effects of the curvature but just due to those of the torsion 
in teleparallelism. 
%%%%%
%%% Unit %%%
We use units of $k_\mathrm{B} = c = \hbar = 1$ and denote the
gravitational constant $8 \pi G$ by
${\kappa}^2 \equiv 8\pi/{M_{\mathrm{Pl}}}^2$
with the Planck mass of $M_{\mathrm{Pl}} = G^{-1/2} = 1.2 \times 
10^{19}$\,\,GeV. 
%%%%%%%%%%%%

%%%%% Structure %%%%%
The Letter is organized as follows. 
In Sec.\ II, we introduce the formulations in teleparallelism 
and first explore the effective 
$F(T)$ gravity in the four-dimensional space-time 
coming from the five-dimensinal Kaluza-Klein (KK) theory.
Next, in Sec.\ III, we examine the RS type-II model and show that 
the $F(T)$ gravity contribution will exist on the brane with 
the four-dimensional flat FLRW space-time. 
In Sec.\ VI, conclusions are finally given. 
%%%%%%%%%%%%%%%%%%%%%

%%%%%%%%%%%%%%%%%%%%%%%%%%%
%%%  Sec. II
%%%%%%%%%%%%%%%%%%%%%%%%%%%
\section{From the Kaluza-Klein (KK) theory}
%\noindent
%%%%%%%%%%%%%%%%%%%%%
%\textit{From the Kaluza-Klein (KK) theory}--- 
%%%%%%%%%%%%%%%%%%%%%

In teleparallelism, 
orthonormal tetrad components $e_A (x^{\mu})$ with $A = 0, 1, 2, 3$ 
are adopted. 
Here, an index $A$ is for the 
tangent space at each point $x^{\mu}$ of the manifold. 
With orthonormal tetrad components, 
the metric is expressed as 
$
g_{\mu\nu}=\eta_{A B} e^A_\mu e^B_\nu 
$ with $\mu$ and $\nu$, where $\mu, \, \nu = 0, 1, 2, 3$, 
coordinate indices on the manifold, and accordingly 
$e_A^\mu$ is equivalent to the tangent vector of the manifold. 
This is called the vierbein. 
%%%
The relation $e^A_\mu e_A^\nu = \delta_\mu^\nu$ 
defines the inverse of the vierbein. 
%%%
The torsion and contorsion tensors are defined as 
$T^\rho_{\verb| |\mu\nu} \equiv 
\Gamma^\rho_{\verb| |\nu\mu} - \Gamma^\rho_{\verb| |\mu\nu} 
= e^\rho_A 
\left( \partial_\mu e^A_\nu - \partial_\nu e^A_\mu \right)$ 
with $\Gamma^\rho_{\verb| |\nu\mu} \equiv e^\rho_A \partial_\mu e^A_\nu$, 
being the Weitzenb\"{o}ck connection without curvature, and 
$K^{\mu\nu}_{\verb|  |\rho} 
\equiv -\left(1/2\right) 
\left(T^{\mu\nu}_{\verb|  |\rho} - T^{\nu \mu}_{\verb|  |\rho} - 
T_\rho^{\verb| |\mu\nu}\right)$ the contortion tensor, respectively. 
The torsion scalar is constructed as 
$T \equiv S_\rho^{\verb| |\mu\nu} T^\rho_{\verb| |\mu\nu} = 
\left(1/4\right)T^{\rho\mu\nu}T_{\rho\mu\nu} + \left(1/2\right) 
T^{\rho\mu\nu}T_{\nu\mu\rho}-T_{\rho\mu}^{\verb|  |\rho}
T^{\nu\mu}_{\verb|  |\nu}$, 
where 
$S_\rho^{\verb| |\mu\nu} \equiv \left(1/2\right)
\left(K^{\mu\nu}_{\verb|  |\rho}+\delta^\mu_\rho \ 
T^{\alpha \nu}_{\verb|  |\alpha}-\delta^\nu_\rho \ 
T^{\alpha \mu}_{\verb|  |\alpha}\right)$ 
is the superpotential. 
Consequently, 
the teleparallel Lagrangian density is 
described by the torsion scalar 
$T$, although the Einstein-Hilbert action is represented by 
the scalar curvature $R$ in general relativity. 
%%%
The modified teleparallel action describing $F(T)$ 
gravity~\cite{Linder:2010py} with matter is 
\begin{equation} 
S= \int d^4x \abs{e} \left( \frac{F(T)}{2{\kappa}^2} 
+{\mathcal{L}}_{\mathrm{M}} \right)\,, 
\label{eq:FT9-15-01}
\end{equation}
where $\abs{e}= \det \left(e^A_\mu \right)=\sqrt{-g}$ with $g$ 
the determinant of the metric $g_{\mu\nu}$ and 
${\mathcal{L}}_{\mathrm{M}}$ is the matter Lagrangian. 
%%%%%
In what follows, we concentrate on the part of gravitation of the action. 
%%%%%

First, we explore the four-dimensional effective $F(T)$ gravity from the 
KK theory. We suppose that the procedure of the KK reduction~\cite{MKK-A-C-F, Overduin:1998pn, F-M} can be applied to the modified teleparallel gravity in the same manner as in general relativity. 
The action of $F(T)$ gravity in the five-dimensional space-time is 
expressed as~\cite{Capozziello:2012zj} 
\begin{eqnarray}
&&
{}^{(5)}S= 
\int d^5 x \left|{}^{(5)}e\right| \frac{
F({}^{(5)}T)}{2 \kappa_5^2}\,,
\label{eq:Add-001} \\
&&
{}^{(5)}T \equiv 
\frac{1}{4} T^{a b c}T_{a b c} + \frac{1}{2} 
T^{a b c}T_{c b a}-T_{a b}^{\verb|  |a}
T^{c b}_{\verb|  |c}\,,
\label{eq:Add-002}
\end{eqnarray} 
where ${}^{(5)}e = \sqrt{{}^{(5)}g}$ with ${}^{(5)}g$ the determinant of 
the metric ${}^{(5)}g_{\mu\nu}$ in the five-dimensional space-time, 
$\kappa_{5}^2 \equiv 8 \pi G_5 = 
\left( {}^{(5)}M_{\mathrm{Pl}} \right)^{-3}$ 
with $G_5$ the gravitational constant and 
$M_{\mathrm{Pl}}^{(5)}$ the Planck mass 
in the five-dimensional space-time. 
Here, the superscript or subscript of $(5)$ or $5$ mean the 
quantities in the five-dimensional space-time. 
%%%%%
In addition, ${}^{(5)}T$ is 
the torsion scalar in the five-dimensional space-time, where 
the Latin indices $a, b, \dots$ run over $0, 1, 2, 3, 5$ 
and ``$5$'' denotes the component of the fifth coordinate. 
The form in Eq.~(\ref{eq:Add-002}) is equivalent to that in the four-dimensional space-time shown above~\cite{Capozziello:2012zj}. 
%%%%%
We now consider the following original KK compactification scenario 
in case of the five-dimensional space-time. 
One of the dimensions of space is compactified to a small circle and 
the four-dimensional space-time is extended infinitely. 
The radius of the fifth dimension is taken to be of order of the Planck 
length in order for the KK effects not to be seen. Thus, the size of 
the circle is so small that phenomena in sufficiently low energies cannot be 
detected~\cite{MKK-A-C-F, Overduin:1998pn, F-M}. 
%%%
Provided that the metric in the five-dimensional space-time is described as 
the following diagonal form 
\begin{equation} 
{}^{(5)}g_{a b} = 
\left( 
\begin{array}{cc} 
g_{\mu\nu} & 0 \\ 
0 & -\phi^2 \\ 
\end{array} 
\right)\,, 
\label{eq:0-1} 
\end{equation}
with $\phi \equiv \varphi/\varphi_{*}$ a homogeneous scalar field depending 
only on time, where $\phi$ is a dimensionless quantity, $\varphi$ is a homogeneous scalar field having a mass dimension and $\varphi_{*}$ is a fiducial value of $\varphi$. 
%%%%%
We represent $\phi^2 = \mathcal{R}^2 \theta^2$, 
where $\mathcal{R}$ is the radius of the compactified space, and 
the orthonormal tetrad components in the one-dimensional compactified space 
is written by the dimensionless 
coordinates $\theta$ such as an angle. 
We also find $\sqrt{{}^{(5)}g}=\sqrt{-g} \mathcal{R} 
\sqrt{\hat{g}}$. 
Here, $\hat{g}$ is the determinant of the metric corresponding to 
the pure geometrical part represented by $\theta$ and 
relevant to the compactified space volume 
$V_{\mathrm{com}} =  \int \hat{g} d \theta$~\cite{F-M}. 
%%%%%
In this case, we take $e^A_a = \mathrm{diag} (1, 1, 1, 1, \phi)$ and 
the $\eta_{a b} = \mathrm{diag} (1, -1, -1, -1, -1)$. 
For the action in the five-dimensional space-time in Eq.~(\ref{eq:Add-001}) 
with Eq.~(\ref{eq:Add-002}), 
by adopting the above expressions of $e^A_a$ and $\eta_{a b}$ 
to analyze ${}^{(5)}S$ and ${}^{(5)}T$, 
the effective action in the four-dimensional space-time 
through the KK compactification mechanism explained above 
can be described as 
\begin{equation}
S_{\mathrm{KK}}^{\mathrm{eff}} = \int d^4x \abs{e} 
\frac{1}{2\kappa^2} \phi 
F(T +\phi^{-2} \partial_{\mu} \phi \partial^{\mu} \phi)\,.
\label{eq:1-2} 
\end{equation}
%
%%%%%
The appearance of $\phi$ on the right-hand side in Eq.~(\ref{eq:1-2}) 
in front of the function $F$ 
comes from the relation $\left|{}^{(5)}e\right| = \phi \abs{e}$ 
due to the KK dimensional reduction. 
Furthermore, the form of ${}^{(5)}T$ is the same as that of $T$, 
and the part of the $0, \dots, 3$ of ${}^{(5)}g_{a b}$ in Eq.~(\ref{eq:0-1}) 
is $g_{\mu\nu}$, i.e., the metric in the four-dimensional space-time. 
Hence, the form of the torsion scalar through the KK dimensional reduction 
to the four-dimensional space-time, 
which is the argument of the function $F$ 
on the right-hand side in Eq.~(\ref{eq:1-2}), 
would consist of $T$ and the other part in terms of $\phi$, 
which is related to the size of the compactified space. 
We note that the form of the function $F$ itself would not 
be changed by the KK dimensional reduction. 
%%%%%
Our KK reduced action in Eq.~(\ref{eq:1-2}) is compatible with the results in Ref.~\cite{Fiorini:2013hva}. 
%%%%%
%%%
Also, we mention that 
the investigations in the case with the non-diagonal form of 
the metric in the five-dimensional space-time have also been executed 
in Ref.~\cite{NDM}. 
%%%
%%%%%
Here, as a simplest example, 
we consider the case of teleparallelism, i.e., $F(T) = T-2\Lambda_4$ in Eq.~(\ref{eq:FT9-15-01}) with $\Lambda_4 (>0)$ a positive cosmological constant in the 
four-dimensional space-time. 
%%%%%
In this case, the action in Eq.~(\ref{eq:1-2}) is similar to 
the one describing the Brans-Dicke theory with the cosmological 
constant. 
If we define a scalar field $\sigma$ as 
$\phi \equiv \xi \sigma^2$ with $\xi = 1/4$, 
we can rewrite the action in Eq.~(\ref{eq:1-2}) into the one 
where the kinetic term of $\sigma$ becomes canonical as~\cite{F-M}
$S_{\mathrm{KK}}^{\mathrm{eff}} |_{F(T)=T-2\Lambda_4} = \int d^4x \abs{e} 
\left(1/\kappa^2 \right) 
\left[ \left(1/8\right) \sigma^2 T + \left(1/2\right) \partial_{\mu} \sigma \partial^{\mu} \sigma - \Lambda_4 \right]$. 
%%%%%
The metric of the flat FLRW universe is written as 
$
ds^2 = dt^2 - a^2(t) \sum_{i=1,2,3}\left(dx^i\right)^2 
$ 
with $a$ the scale factor and $H \equiv \dot{a}/a$ the Hubble parameter, 
where the dot denotes the time derivative of $\partial/\partial t$. 
For this space-time, we have 
$g_{\mu \nu}= \mathrm{diag} (1, -a^2, -a^2, -a^2)$ and 
$e^A_\mu = \mathrm{diag} (1,a,a,a)$. 
These expressions lead to the relation $T=-6H^2$. 
%%%%%
In this background, 
the gravitational field equations read 
$\left(1/2\right) \dot{\sigma}^2 
-\left(3/4\right) H^2 \sigma^2 + \Lambda_4 = 0$ and 
$\dot{\sigma}^2 + H \sigma \dot{\sigma} 
+\left(1/2\right) \dot{H} \sigma^2 = 0$~\cite{Geng:2011aj}. 
%%%
Furthermore, the equation of motion of $\sigma$ becomes 
$\ddot{\sigma} + 3H\dot{\sigma} + \left(3/2\right) H^2 \sigma = 0$. 
In deriving these equations, we have used the relation $T=-6H^2$. 
%%%%%%%%
By combining the above gravitational field equations, we have 
$\left(3/2\right) H^2 \sigma^2 -2\Lambda_4 + H \sigma \dot{\sigma} 
+ \left(1/2\right) \dot{H} \sigma^2 = 0$. 
Hence, we can obtain a solution for this equation as 
$H = H_{\mathrm{inf}} = \mathrm{constant} (>0)$, 
which corresponds to the Hubble parameter at the inflationary stage, 
and $\sigma = b_1 \left(t/t_1\right) + b_2$, where 
$b_1$ is a constant and $b_2 (>0)$ a positive one, and 
$t_1$ denotes a time. 
In the limit $t \to 0$, we can acquire an approximate 
expression as 
$H_{\mathrm{inf}} \approx \left(2/b_2\right) \sqrt{\Lambda_4/3}$ 
and $\sigma \approx b_2$. 
Furthermore, with the equation of motion of $\sigma$, 
for $t \to 0$, 
we find $b_1 \approx -\left(1/2\right) b_2 H_{\mathrm{inf}} t_1 
\approx -\sqrt{\Lambda_4/3} t_1$. 
As a result, when $t \to 0$, an exponential inflation 
with the scale factor 
$a \approx \bar{a} \exp \left( H_{\mathrm{inf}} t \right)$, 
where $\bar{a} (>0)$ is a positive constant, 
can be realized approximately. 
%%%%%%%%
It is significant to emphasize that the contribution of the effect of 
the KK compactification, namely, the role of extra dimensions, is to 
lead to the scalar field $\phi$ in the gravitational field equations. 
%%%

%%%%%%%%%%%%%%%%%%%%%%%%%%%
%%%  Sec. III
%%%%%%%%%%%%%%%%%%%%%%%%%%%
\section{From the Randall-Sundrum (RS) theory}
%\noindent
%%%%%%%%%%%%%%%%%%%%%
%\textit{From the Randall-Sundrum (RS) theory}--- 
%%%%%%%%%%%%%%%%%%%%%

Next, we explore the four-dimensional effective $F(T)$ gravity from the 
RS theory with the procedure in Ref.~\cite{F-M}. 
In the RS type-I model~\cite{Randall:1999ee}, there are a positive tension brane at $y=0$ and a negative one 
at $y=s$, where $y$ is the fifth direction. 
Suppose that the metric describing 
the five-dimensional space-time is given by 
$ds^2 = \e^{-2|y|/l} g_{\mu\nu} (x) dx^{\mu} dx^{\nu} + dy^2
$ with $l=\sqrt{-6/\Lambda_5}$, where $\e^{-2|y|/l}$ 
is the warp factor and 
$\Lambda_5 (< 0)$ is the negative cosmological constant in the bulk. 
%%%%%
It is known that for the RS type-I model, the effective gravity theory in four-dimensions is the Brans-Dicke (BD) theory with the BD parameter $\omega_{\mathrm{BD}} =\left(3/2\right) \left(\e^{\pm s/l}-1\right)$, where the sign ($\pm$) 
corresponds to that of the brane tension~\cite{Garriga:1999yh}. 
%%%%%

On the other hand, in the RS type-II model~\cite{Randall:1999vf}, 
there is only one brane with the positive tension floating 
in the AdS bulk space and hence 
the negative-tension brane does not exist. This configuration can be 
realized by the RS type-I model~\cite{Randall:1999ee} 
with two branes in the limit $s \to \infty$. 
%%%%%
We start with the equation in the five-dimensional space-time 
with the brane whose tension is a positive constant. 
We consider that 
the vacuum solution in the five-dimensional space-time is AdS one, 
and that 
the brane configuration is consistent with the equation in the five-dimensional space-time. 
This implies that 
the brane configuration with a positive constant tension 
connecting two vacuum solutions in the five-dimensional space-time, 
namely, 
the condition of the 
configuration is nothing but the equation for the brane.
%%%%%
%%% 
In Ref.~\cite{Nozari:2012qi}, using 
the analysis in Ref.~\cite{Shiromizu:1999wj}, 
the RS type-II model in teleparallelism has been considered.
%%%%%
The procedure is as follows. 
(i) 
The corresponding Gauss-Codazzi equations in teleparallelism, 
namely, the induced equations on the brane, 
is examined by using the projection vierbein of the five-dimensional 
space-time quantities into the four-dimensional space-time brane. 
(ii) 
The Israel's junction conditions to connect the left-side and right-side 
bulk spaces sandwiching the brane are investigated. 
The first junction condition is that the vierbeins induced on the brane 
from the left-side and right-side of the brane should be the same 
with each other. 
Moreover, the second junction condition is that the difference 
of the superpotential between the left-side and right-side of the brane 
comes from the energy-momentum tensor of matter, 
which is confined in the brane. 
(iii) 
Provided that there exists 
$Z_2$ symmetry, i.e., $y \leftrightarrow -y$, 
in the five-dimensional space-time, the quantities on the left and right sides 
of the brane are explored. 
%%%%%
The difference between the scalar curvature and the torsion scalar is a total derivative of the torsion tensor~\cite{Capozziello:2012zj, W-C}. 
This may affect the boundary. 
It has been shown that 
in comparison with the gravitational field equations 
in general relativity~\cite{Shiromizu:1999wj, F-RSII}, 
the induced gravitational field equations on the brane have new terms, 
which comes from the additional degrees of freedom in teleparallelism. 
%%%
These extra terms correspond to the projection on the brane 
of the vector portion of the torsion tensor in the bulk. 
%%%%%

Through the procedure explained above, 
we find that for $F(T)$ gravity, 
in the flat FLRW background the Friedmann equation on the brane is 
given by
\begin{eqnarray} 
&&
H^2 \frac{d F(T)}{dT} = -\frac{1}{12} \left[ F(T) - 4 \Lambda 
-2 \kappa^2 \rho_{\mathrm{M}} 
\right. 
\nonumber \\
&& 
\hspace{28mm}
\left. 
- \left(\frac{\kappa_5^2}{2}\right)^2 
\mathcal{Q} 
\rho_{\mathrm{M}}^2
\right]\,,
\label{eq:3-1}
\end{eqnarray}
with 
$\mathcal{Q} \equiv \left(11-60w_{\mathrm{M}} +93 w_{\mathrm{M}}^2 \right)/4$. 
%%%
We note that $\mathcal{Q}$ includes the contributions from teleparallelism, 
which do not exist in general relativity~\cite{Nozari:2012qi}. 
%%%
Here, $w_{\mathrm{M}} \equiv P_{\mathrm{M}}/\rho_{\mathrm{M}}$ with 
the energy density 
$\rho_{\mathrm{M}}$ and pressure $P_{\mathrm{M}}$ of matter, 
assumed to be a perfect fluid, 
is the equation of state parameter for matter confined to the brane, 
the effective cosmological constant in the brane is 
$\Lambda \equiv \Lambda_5 + \left(\kappa_5^2/2\right)^2 \lambda^2$ 
with $\lambda (> 0)$ the tension of the brane 
and $G = \left[1/\left(3\pi \right) \right] \left(\kappa_5^2/2\right)^2 
\lambda$. 
Clearly, the significant contributions from the fifth dimension 
to the Friedmann equation on the brane are the second term 
and the fourth term 
proportional to $\rho_{\mathrm{M}}^2$ on the right-hand side 
in Eq.~(\ref{eq:3-1}). 
%%%
Furthermore, 
the function of $F(T)$ induced on the brane would be considered to be 
the same as that in the five-dimensional space-time. 
%%%
In the dark energy dominated stage, 
the energy density of non-relativistic matters with $w_{\mathrm{M}} =0$, 
i.e., cold dark matter and baryon, 
is so much smaller than that of the cosmological constant 
that the third and fourth terms 
on the right-hand side can be neglected. 
For teleparallelism with the cosmological constant in the five-dimensional space-time, 
$F(T) = T - 2 \Lambda_5$ in Eq.~(\ref{eq:3-1}), 
we obtain an approximate de Sitter solution on the brane 
$H = H_{\mathrm{DE}} = 
\sqrt{\Lambda_5+\kappa_5^4\lambda^2/6}
= \mathrm{constant}$ 
and $a(t) = a_{\mathrm{DE}} \exp \left( H_{\mathrm{DE}} t \right)$ with 
$a_{\mathrm{DE}} (>0)$ a constant, where we have used $T=-6H^2$. 
Therefore, for the late time cosmic acceleration can be realized. 
%%%%%
We mention that 
for $F(T)=T$, $\Lambda=0$ and ${\mathcal Q}=8/3$ 
realizing if $w_{\mathrm{M}} = -5.5 \times 10^{-3}$, 
we find $H^2 = \left(\kappa^2/3\right) \rho_{\mathrm{M}} 
\left[1 + \rho_{\mathrm{M}}/\left(2 \lambda \right) \right]$, 
which is equivalent to the Friedmann equation 
in the brane world scenario~\cite{Astashenok:2013rp}. 
%%%%%
Moreover, 
for a power-law model such as $F(T) = T^2/\bar{M}^2 +\alpha \Lambda_5$ 
in Eq.~(\ref{eq:3-1}), 
where $\bar{M}$ is a mass scale and $\alpha$ is a constant, 
we find a similar approximate de Sitter solution 
$H = H_{\mathrm{DE}} = 
\left[\left(\bar{M}^2/108\right) \mathcal{J} \right]^{1/4} 
= \mathrm{constant}$ 
with $\mathcal{J} \equiv 
\left(\alpha-4\right) \Lambda_5 
-4 \left(\kappa_5^2/2\right)^2 \lambda^2$, 
%%%
where $\mathcal{J} (\geq 0)$ has to be larger than or equal to zero, 
so that this can lead to a constraint on $\alpha$ as 
$\alpha \geq 4 + \left(\kappa_5^2 \lambda^2\right)/\Lambda_5$. 
%%%
Here, we have used an approximation that 
on the right-hand side of Eq.~(\ref{eq:3-1}), 
the first and second terms, which corresponds to 
the components of dark energy, are much larger than 
the third and fourth terms proportional to $\rho_{\mathrm{M}}$ 
and $\rho_{\mathrm{M}}^2$, respectively. 
This approximation can be appropriate when the universe is considered to be 
the dark energy (sufficiently) dominated stage and thus the energy density of 
non-relativistic matters $\rho_{\mathrm{M}}$ can be negligible in comparison 
with the dark energy density. 
In addition, we note that in deriving the above solution, we have used 
the relation $T=-6 H^2$, and that as a result, both the left-hand side of Eq.~(\ref{eq:3-1}) and the first term of the right-hand side are proportional to 
$H^4$. 
%%%
It is emphasized that 
the generic formulation for the gravitational field equation on the 
brane in teleparallel gravity has been derived in Ref.~\cite{Nozari:2012qi}. 
On the other hand, the new ingredients obtained in this paper 
would be considered to describe the Friedmann equation~(6) 
in the flat FLRW space-time 
and to acquire the solutions to realize the current accelerated expansion of 
the universe for two concrete $F(T)$ models. 
%%%%%
It should also be cautioned that 
the condition on $F(T)$ gravity for 
the AdS configuration in the bulk to be 
realized has to be shown in future work. 
%%%%%
In addition, 
four-dimensional bouncing $F(T)$ cosmologies~\cite{deHaro:2012zt} 
unifying inflation with the late-time cosmic acceleration 
due to dark energy have been discussed.  
Such cosmologies may be reconstructed also 
in $F(T)$ gravity from the RS brane world scenario. 
%%%%%

%%%%%%%%%%%%%%%%%%%
%%%  Sec. IV
%%%%%%%%%%%%%%%%%%%
\section{Conclusions}
%%%%%%%%%%%%%%%%%%%%%
%\textit{Summary}--- 
%%%%%%%%%%%%%%%%%%%%%

We have studied the four-dimensional effective $F(T)$ gravity 
coming from the higher-dimensional KK and RS space-time theories. 
With the KK reduction from the five-dimensional space-time 
to the four dimensions, 
we have built the four-dimensional effective theory 
of $F(T)$ gravity coupling to a scalar field. 
Moreover, for the RS type-II model consisting of 
the five-dimensional AdS space-time and the four-dimensional FLRW brane, 
we have also shown that the contribution of $F(T)$ gravity 
appears on the four-dimensional FLRW brane. 
Furthermore, it has been verified that 
inflation or the late time cosmic accelerated expansion 
can occur only through the effect of the torsion 
without that of the curvature. 
Thus, these models can be regarded as the KK and RS models 
constructed by not the curvature effect but only the torsion one 
in teleparallelism. 

%%%%%
What it has been executed in this Letter is to explicitly demonstrate that 
in the four-dimensional effective $F(T)$ gravity theories obtained 
by the KK reduction from the five-dimensional space-time 
and those on the four-dimensional FLRW brane in 
the RS type-II model, inflation in the early universe and 
and the accelerated expansion in the late time universe 
can be realized, respectively, owing to the effect of the torsion 
of the space-time and not the curvature effect. 
Indeed, this is the first work on the concrete cosmological solutions 
to describe the cosmic accelerated expansion of the KK and RS models 
in $F(T)$ gravity. 
These results may imply that phenomenological $F(T)$ gravity models 
in the four-dimensional space-time 
can be derived from more fundamental theories. 
Since $F(T)$ gravity models can lead to the current accelerated expansion 
of the universe, namely, a resolution of the dark energy problem, 
this study may present us a clue to explore the origin of extensions of 
gravity from general relativity including $F(T)$ gravity. 
%%%%%

Finally, it should be remarked that the observational constraints on 
the derivative of $F(T)$ with respect to $T$ until the fifth order 
have been presented in Ref.~\cite{Capozziello:2011hj} with cosmographic 
parameters acquired from the observational data of Supernovae Ia and the baryon acoustic oscillations. In this Letter, as concrete examples of $F(T)$ gravity models, we have considered $F(T) = T$ plus an effective cosmological constant 
and $F(T) = T^2/\bar{M}^2 $ plus a constant term corresponding to 
an effective cosmological constant. 
These two models can be consistent with the results obtained 
in Ref.~\cite{Capozziello:2011hj}.

%%%%%%%%%%%%%%%%%%%%%%%%
%%%  Acknowledgments
%%%%%%%%%%%%%%%%%%%%%%%%
\section*{Acknowledgments}
%%%%%%%%%%%%%%%%%%%%%
%\textit{Acknowledgments}-- 
%%%%%%%%%%%%%%%%%%%%%

We sincerely thank Professor Chao-Qiang Geng 
for very kind hospitality in Taiwan (K.B. and S.N.). 
We also appreciate important comments of Professor Jaume de Haro 
very much. 
The work is supported in part by 
Global COE Program of Nagoya University (G07) 
provided by the Ministry of Education, Culture, Sports, Science \&
Technology and 
by the JSPS Grant-in-Aid for 
Young Scientists (B) \# 25800136 (K.B.);\ 
that for Scientific Research 
(S) \# 22224003 and (C) \# 23540296 (S.N.);\ 
and 
MINECO (Spain), FIS2010-15640, 
AGAUR (Generalitat de Ca\-ta\-lu\-nya), contract 2009SGR-345, 
and project 2.1839.2011 
of Min. of Education and Science (Russia) (S.D.O.).

%%%%%%%%%%%%%%%%%%%%%%%%%%%%%%%%%
%% thebibliography environment
%%%%%%%%%%%%%%%%%%%%%%%%%%%%%%%%%

\end{document}